\documentstyle[12pt]{article}
\begin{document}
\title{{\small\centerline{January 1995\hfill IIA-TH/95-4}}
\medskip
{\bf Towards a physical interpretation of the angular momentum parameter
associated with the Kerr metric}}
\author{\bf Sujan Sengupta\thanks{e-mail : sujan@iiap.ernet.in}\\
\normalsize \em Indian Institute Of Astrophysics, Bangalore 560 034, INDIA}
\date{}
\maketitle
\begin{abstract} \normalsize\noindent
The role of the parameter `$a$' associated with the Kerr metric
which represents the angular momentum per unit mass of a
rotating  massive object has been converted into determining the
rotationally induced quadrapole electric field outside a rotating massive
object with external dipole magnetic field. A comparison of the
result with that of the Newtonian case implies that the
parameter `$a$'  represents the angular momentum per unit mass
of a rotating hollow object.
\end{abstract}

\newpage
The first exact solution of Einstein equations to be found which
could represent the exterior field of a stationary,
axisymmetric, asymptotically flat field outside a bounded
rotating massive object is the Kerr metric which in
Boyer-Lindquist co-ordinates may be written as
          \begin{eqnarray}
ds^{2}&=&-(1-2mr/\Sigma)c^{2}dt^{2}-\frac{4mra}{\Sigma}\sin^{2}\theta
cdtd\phi +(\Sigma/\Delta)dr^{2}+\Sigma d\theta^{2}+ \nonumber \\
& & (B/\Sigma)\sin^{2}\theta d\phi^{2}
          \end{eqnarray}
where
$$\Sigma=(r^{2}+a^{2}\cos^{2}\theta), \Delta=(r^{2}+a^{2}-2mr),
B=(r^{2}+a^{2})^{2} -\Delta a^{2}\sin^{2}\theta,$$
$$m=MG/c^{2},
a=\frac{GJ}{mc^{3}}$$
and $J$ is the total angular momentum.

There have been several attempts to find an interior solution
which matches smoothly onto the Kerr solution, that is, an
interior solution with a boundary at which the metric components
and their derivatives will be continuous at the boundary. No
such solution has yet been found. The search for a physical
source that produces the Kerr field has profound significance in
the sense that in that situation the Kerr field would resemble
the case of the spherically symmetric Schwarzschild metric which
can represent both a black hole and the external field due to
matter.

The two parameter Kerr metric reduces to the `physically
acceptable' Schwarzschild metric when one of the parameters
`$a$' interpreted from its asymptotic behaviour as the total
angular momentum of the object \cite{1} tends to zero. Therefore
it is this extra parameter which requires a physical
interpretation in order to obtain a physical picture of the Kerr
solution.

The problem of physically interpreting the parameter `$a$' is
made difficult by the lack of a classical analogue of the Kerr
metric because in classical gravitational theory the field of an
axially symmetric body is independent of its rotational motion,
unlike the situation in relativity.

This prohibitive condition can be tackled by converting the role
played by the parameter `$a$' from contributing to the spacetime
geometry into the determination of another physical quantity
which does have a classical analogue. In this letter the
rotationally induced electric field of an external dipole
magnetic field superposed in Kerr background geometry is chosen
to interpret the physical meaning of `$a$' from its classical
counterpart.

I consider a massive rotating object (other than a black hole)
of total mass $M$ and total angular momentum $J$. Further I assume
that the object has an aligned dipole external magnetic field
with arbitrary dipole moment $\mu$. The spacetime geometry
external to this object is described by the Kerr metric with the
additional assumption that the magnetic field is not so strong
that it affects the geometry significantly.

Several authors have obtained the solution for stationary
axisymmetric electromagnetic field around a rotating massive
object in source free regions of spacetime by using Teukolsky's
perturbation equations in Newman-Penrose formalism \cite{2,3,4}.
I shall use the expression for the vector potential as given in
\cite{3} restricting myself to the case of a dipole magnetic
field with no electrostatic charge (Q=0) in a source free region
of spacetime which is given by
$$A_{i}=(A_{t},0,0,A_{\phi})$$
with
        \begin{eqnarray}
A_{t}&=&(\frac{a\beta^{i}_{1}}{\Sigma})\{[r(r-m)+(a^{2}-mr)
\cos^{2}\theta]+\frac{1}{2\gamma}\ln(\frac{r-m+\gamma}{r-m-
\gamma})- \nonumber \\
& &(r-m\cos^{2}\theta)\}-(\frac{\beta^{r}_{1}\cos\theta}{\Sigma})
\{(a^{2}\sin^{2}\theta-r^{2}+mr)+ \nonumber \\
& &[(r^{3}-2mr^{2}+ma^{2})+(r-m)a^{2}\cos^{2}\theta]\frac{1}{2\gamma}
\ln(\frac{r-m+\gamma}{r-m-\gamma})\}
      \end{eqnarray}
      \begin{eqnarray}
A_{\phi}&=&(\frac{\beta^{i}_{1}\sin^{2}\theta}{2\Sigma})\{r(r^{2}
+mr+2a^{2})+(r-m)a^{2}\cos^{2}\theta-[r(r^{3}- \nonumber \\
& &2ma^{2}+a^{2}r)+\Delta a^{2}\cos^{2}\theta]
\frac{1}{2\gamma}\ln(\frac{r-m+\gamma}{r-m-\gamma})\}+ \nonumber \\
& &(\frac{a\beta_{1}^{r}\sin^{2}\theta\cos\theta}{\Sigma})\{(a^{2}+mr)
+\frac{m(a^{2}-r^{2})}{2\gamma}\ln(\frac{r-m+\gamma}{r-m-\gamma})\}
   \end{eqnarray}
with
$\gamma=(m^{2}-a^{2})^{1/2},
\beta_{1}^{r}$ and $\beta^{i}_{1}$ being constants.
In the above expressions higher multipoles [$\geq O(a^2)$] are
assumed to be negligible.

Since, in the absence of rotation $(a=0)$ there is no electric
field and $A_{t}=0$, one immediately gets the constant
$\beta^{r}_{1}=0$. In order to determine $\beta^{i}_{1}$, I
consider the boundary condition, that the Kerr geometry is
asymptotically flat and thus as $r\rightarrow\infty$ the
magnetic field and the electric field should have the structure
of dipole and quadrapole fields on flat spacetime respectively.

Accordingly one  gets
$$\beta^{i}_{1}=\mp(3\mu/2\gamma^{2})$$
depending on whether the dipole moment $\mu$ is parallel or
antiparallel to the rotation axis. However in the present study
I shall consider only the case with
$\beta^{i}_{1}=-3\mu/2\gamma^{2}$. Hence I get the components of
the vector potential as \cite{5}
       \begin{eqnarray}
A_{t}&=&-\frac{3a\mu}{2\gamma^{2}\Sigma}\{[r(r-m)+(a^{2}-mr)
\cos^{2}\theta]\frac{1}{2\gamma}\ln(\frac{r-m+\gamma}{r-m-\gamma})
- \nonumber \\
& &(r-m\cos^{2}\theta)\}
   \end{eqnarray}
   \begin{eqnarray}
A_{\phi}&=&-\frac{3\mu\sin^{2}\theta}{4\gamma^{2}\Sigma}\{(r-m)
a^{2}\cos^{2}\theta+r(r^{2}+mr+2a^{2})- \nonumber \\
& &[r(r^{3}-2ma^{2}+a^{2}r)+\Delta a^{2}\cos^{2}\theta]
\frac{1}{2\gamma}\ln(\frac{r-m+\gamma}{r-m-\gamma})\}
    \end{eqnarray}

Using the definition $F_{ij}=(A_{j,i}-A_{i,j})$, I obtain the
components of the quadrapole electric field (induced) as well as
the components of the dipole magnetic field in Kerr geometry as:
      \begin{eqnarray}
F_{rt}&=&-\frac{3a\mu}{2\gamma^{2}(r^{2}+a^{2}\cos^{2}\theta)}
\{\frac{1}{2\gamma}(2r-m-m\cos^{2}\theta)\ln(\frac{r-m+\gamma}{r-m-
\gamma})-\nonumber \\
& &\frac{1}{(r-m+\gamma)(r-m-\gamma)}[r(r-m)+(a^{2}-mr)\cos^{2}\theta]
-1\}+ \nonumber \\
& &\frac{3a\mu r}{\gamma^{2}(r^{2}+a^{2}\cos^{2}\theta)^{2}}
\{-r+m\cos^{2}\theta+\frac{1}{2\gamma}[r(r-m)+(a^{2}-mr)
\cos^{2}\theta]\times \nonumber \\
& &\ln(\frac{r-m+\gamma}{r-m-\gamma})\}
    \end{eqnarray}
     \begin{eqnarray}
F_{\theta t}&=&-\frac{3a^{3}\mu\cos\theta\sin\theta}{\gamma^{2}(r^{2}
+a^{2}\cos^{2}\theta)^{2}}\{m\cos^{2}\theta+
\frac{1}{2\gamma}[r(r-m)+(a^{2}-mr)\cos^{2}\theta]\times \nonumber \\
& &\ln(\frac{r-m+\gamma}{r-m-\gamma})-r\}+\frac{3a\mu}{2\gamma^{2}
(r^{2}+a^{2}\cos^{2}\theta)}\{2m\cos\theta\sin\theta+ \nonumber \\
& &\frac{1}{\gamma}[(a^{2}-mr)\cos\theta\sin\theta\ln(\frac{r-m+
\gamma}{r-m-\gamma})]\}
     \end{eqnarray}
     \begin{eqnarray}
F_{r\phi}&=&-\frac{3\mu\sin^{2}\theta}{4\gamma^{2}(r^{2}+a^{2}\cos^{2}
\theta)}\{r^{2}+mr+2a^{2}+r(m+2r)+a^{2}\cos^{2}\theta+ \nonumber \\
& &\frac{1}{(r-m+\gamma)(r-m-\gamma)}[r(r^{3}-2a^{2}m+a^{2}r)+a^{2}
(r^{2}-2mr+a^{2})\cos^{2}\theta]- \nonumber \\
& &\frac{1}{2\gamma}[r^{3}-2a^{2}m+a^{2}r+r(a^{2}+3r^{2})+a^{2}
(2r-2m)\cos^{2}\theta]\ln(\frac{r-m+\gamma}{r-m-\gamma})\}+ \nonumber \\
& &\frac{3\mu r\sin^{2}\theta}{2\gamma^{2}(r^{2}+a^{2}\cos^{2}
\theta)^{2}}\{r(r^{2}+mr+2a^{2})+a^{2}(r-m)\cos^{2}\theta- \nonumber \\
& &\frac{1}{2\gamma}[r(r^{3}-2a^{2}m+a^{2}r)+a^{2}(r^{2}-2mr+a^{2})
\cos^{2}\theta]\ln(\frac{r-m+\gamma}{r-m-\gamma})\}
      \end{eqnarray}
        \begin{eqnarray}
F_{\theta \phi}&=&-\frac{3\mu\cos\theta\sin\theta}{2\gamma^{2}
(r^{2}+a^{2}\cos^{2}\theta)}\{r(r^{2}+mr+2a^{2})+a^{2}(r-m)
\cos^{2}\theta- \nonumber \\
& &\frac{1}{2\gamma}[r(r^{3}+a^{2}r-2a^{2}m)+a^{2}(r^{2}-2mr
+a^{2})\cos^{2}\theta]\ln(\frac{r-m+\gamma}{r-m-\gamma})\}- \nonumber \\
& &\frac{3a^{2}\mu\cos\theta\sin^{3}\theta}{2\gamma^{2}(r^{2}+
a^{2}\cos^{2}\theta)^{2}}\{r(r^{2}+mr+2a^{2})+a^{2}(r-m)\cos^{2}\theta-
\frac{1}{2\gamma}[r(r^{3}+ \nonumber \\
& &a^{2}r-2a^{2}m)+
a^{2}(r^{2}-2mr+a^{2})\cos^{2}\theta]\ln(\frac{r-m+\gamma}{r-
m-\gamma})\}- \nonumber \\
& &\frac{3\mu a^{2}\sin^{2}\theta}{4\gamma^{2}(r^{2}+a^{2}
\cos^{2}\theta)}\{\frac{\sin\theta\cos\theta}{\gamma}(r^{2}-
2mr+a^{2})\ln(\frac{r-m+\gamma}{r-m-\gamma})- \nonumber \\
& &2(r-m)\cos\theta\sin\theta\}
     \end{eqnarray}

Using the orthonormal tetrad components for the Local Lorentz
frame (LLF) as well as Locally non-rotating frame (LNRF) for the
Kerr metric \cite{1,6} one obtains the $r$ component of the
induced quadrapole electric field in LNRF as
      \begin{eqnarray}
E_{r}&=&-\frac{3a\mu B^{1/2}}{2\gamma^{2}\Sigma^{2}}\{\frac{1}{2\gamma}
(2r-m-m\cos^{2}\theta)\ln(\frac{r-m+\gamma}{r-m-\gamma})- \nonumber \\
& &\frac{1}{(r-m+\gamma)(r-m-\gamma)}[r(r-m)+(a^{2}-mr)\cos^{2}\theta]-1\}
+\frac{3a\mu rB^{1/2}}{\gamma^{2}\Sigma^{3}}\times \nonumber \\
& &\{-r+m\cos^{2}\theta+\frac{1}{2\gamma}[r(r-m)+(a^{2}-mr)
\cos^{2}\theta]\ln(\frac{r-m+\gamma}{r-m-\gamma})\}- \nonumber \\
& &\frac{3a\mu mr\sin^{2}\theta}{2\gamma^{2}B^{1/2}\Sigma^{2}}\{r^{2}+
mr+2a^{2}+r(m+2r)+a^{2}\cos^{2}\theta+\frac{1}{(r-m+\gamma)
(r-m-\gamma)}\times \nonumber \\
& &[r(r^{3}-2a^{2}m+a^{2}r)+a^{2}(r^{2}-2mr+a^{2})\cos^{2}\theta]-
\frac{1}{2\gamma}[4r^{3}-2a^{2}m+2a^{2}r+ \nonumber \\
& &2a^{2}\cos^{2}\theta(r-m)]\ln(\frac{r-m+\gamma}{r-m-\gamma})\}+
\frac{3a\mu mr^{2}\sin^{2}\theta}{\gamma^{2}\Sigma^{3}B^{1/2}}
\{r(r^{2}+mr+2a^{2})+ \nonumber \\
& &a^{2}(r-m)\cos^{2}\theta-
\frac{1}{2\gamma}[r(r^{3}-2a^{2}m+a^{2}r)+a^{2}(r^{2}
-2mr+a^{2})\cos^{2}\theta]\times \nonumber \\
& &ln(\frac{r-m+\gamma}{r-m-\gamma})\}
       \end{eqnarray}

Similarly, in the LLF
     \begin{eqnarray}
E_{r}&=&-\frac{3a\mu(r^{2}+a^{2})}{2\gamma^{2}\Sigma^{2}}
\{\frac{1}{2\gamma}(2r-m-m\cos^{2}\theta)
\ln(\frac{r-m+\gamma}{r-m-\gamma})- \nonumber \\
& &\frac{1}{(r-m+\gamma)(r-m-\gamma)}[r(r-m)+(a^{2}-mr)\cos^{2}\theta]
-1\}+\frac{3a\mu r(r^{2}+a^{2})}{\gamma^{2}\Sigma^{3}}\times \nonumber \\
& &\{-r+m\cos^{2}\theta+\frac{1}{2\gamma}[r(r-m)+(a^{2}-mr)
\cos^{2}\theta]\ln(\frac{r-m+\gamma}{r-m-\gamma})\}- \nonumber \\
& &\frac{3a\mu\Delta^{1/2}\sin^{3}\theta}{4\gamma^{2}\Sigma^{2}}
\{r^{2}+mr+2a^{2}+r(m+2r)+a^{2}\cos^{2}\theta+ \nonumber \\
& &\frac{1}{(r-m+\gamma)(r-m-\gamma)}
[r(r^{3}-2a^{2}m+a^{2}r)+a^{2}(r^{2}-2mr+a^{2})\cos^{2}\theta]- \nonumber \\
& &\frac{1}{2\gamma}[4r^{3}-2a^{2}m+2a^{2}r+
2a^{2}\cos^{2}\theta(r-m)]\ln(\frac{r-m+\gamma}{r-m-\gamma})\}+ \nonumber \\
& &\frac{3a\mu\Delta^{1/2}\sin^{3}\theta}{2\gamma^{2}\Sigma^{3}}
\{r(r^{2}+mr+2a^{2})+a^{2}(r-m)\cos^{2}\theta-
\frac{1}{2\gamma}[r(r^{3}-2a^{2}m+a^{2}r)+ \nonumber \\
& &a^{2}(r^{2}-2mr+a^{2})\cos^{2}\theta]\ln(\frac{r-m+\gamma}{r-m-\gamma})\}
     \end{eqnarray}

Now for arbitrarily slow rotation $(a<<m)$ equation (10) reduces to
     \begin{eqnarray}
E_{r}&=&-\frac{3a\mu}{2m^{2}r^{2}}\{-\frac{1}{2m}(2r-m-m\cos^{2}
\theta)\ln(1-2m/r)- \nonumber \\
& &\frac{1}{(r-2m)}[r-m-m\cos^{2}\theta]-1\}+\frac{3a\mu}{m^{2}r^{3}}
\{-r+m\cos^{2}\theta- \nonumber \\
& &\frac{1}{2m}[r(r-m)-mr\cos^{2}\theta]\ln(1-2m/r)\}-
\frac{3a\mu\sin^{2}\theta}{2mr^{5}}\{3r^{2}+ \nonumber \\
& &2mr+\frac{r^{3}}{r-2m}+\frac{2r^{3}}{m}\ln(1-2m/r)\}+
\frac{3a\mu\sin^{2}\theta}{mr^{6}}\{r^{3}+ \nonumber \\
& &mr^{2}+\frac{r^{4}}{2m}\ln(1-2m/r)\}
       \end{eqnarray}

Similarly, equation (11) reduces to
          \begin{eqnarray}
E_{r}&=&-\frac{3a\mu}{2m^{2}r^{2}}\{-\frac{1}{2m}(2r-m-m\cos^{2}
\theta)\ln(1-2m/r)- \nonumber \\
& &\frac{1}{(r-2m)}[r-m-m\cos^{2}\theta]-1\}+\frac{3a\mu}{m^{2}r^{3}}
\{-r+m\cos^{2}\theta- \nonumber \\
& &\frac{1}{2m}[r(r-m)-mr\cos^{2}\theta]\ln(1-2m/r)\}+\frac{3a\mu
\sin^{3}\theta(r^{2}-2mr)^{1/2}}{4m^{2}r^{4}}\{3r^{2}+ \nonumber \\
& &2mr+\frac{r^{3}}{r-2m}+\frac{2r^{3}}{m}\ln(1-2m/r)\}-\frac{3a\mu
\sin^{3}\theta(r^{2}-2mr)^{1/2}}{2m^{2}r^{6}}\{r^{3}+ \nonumber \\
& &mr^{2}+\frac{r^{4}}{2m}\ln(1-2m/r)\}
       \end{eqnarray}

Along the axis of symmetry $(\theta=0)$, both the equations (12)
and (13) reduces to
      \begin{eqnarray}
E_{r}=\frac{3a\mu}{2m^{2}r^{2}}[\ln(1-\frac{2m}{r})+\frac{2m}{r}]
       \end{eqnarray}

In the above expression the parameter $a$ plays the role in
determining the induced qudrapole electric field only and its
contribution to the spacetime geometry is negligible. Hence the
spacetime geometry surrounding the region is similar to the
spherically symmetric Schwarzschild external geometry.

Asymptotically when $r\rightarrow\infty$ one obtains from
equation (14)
    \begin{eqnarray}
E_{r}=-\frac{3a\mu}{r^{4}}
      \end{eqnarray}

If $R$ is the radius of the sphere then the
electric field at the pole is
        \begin{eqnarray}
E_{r}=-\frac{3a\mu}{R^{4}}
        \end{eqnarray}

Under analogus circumstances the induced quadrapole electric
field can be obtained for the Newtonian case in the following
way:

Consider a thin layer adjacent to and co-rotating with a
spherical object of radius $R$ which if nonrotating, would have a
dipolar external magnetic field given by
        \begin{eqnarray}
{\bf B}=\frac{2\mu}{r^{3}}(\cos\theta,\frac{1}{2}\sin\theta,0)
        \end{eqnarray}
and is continuous across the layer.  The above
expression can be obtained from equations (8) and (9) by
transfering the quantities into LLF, taking $a=0$ and
$r\rightarrow\infty$. Suppose the electric conductivity of the
material within the layer is effectively infinite so that Ohm's
law takes the form within the layer as
     \begin{eqnarray}
{\bf E^{l}}+\frac{\bf \Omega\times r}{c}\times{\bf B}=0
      \end{eqnarray}
where ${\bf E^{l}}$ represents the electric field
within the layer and ${\bf\Omega}$ is the angular velocity
vector of the object.

In the absence of any source, both the normal and the tangential
components of ${\bf B}$ are continuous across the layer so that
equation (18) gives
      \begin{eqnarray}
{\bf E^{l}}=\frac{2\mu\Omega}{cr^{2}}(\frac{1}{2}\sin^{2}\theta,
-\sin\theta\cos\theta,0)
     \end{eqnarray}

Since the tangential component of ${\bf E}$ is
continuous across the layer, so just outside it equation (19)
implies
      \begin{eqnarray}
E^{out}_{\theta}=\frac{\partial}{\partial\theta}
(\frac{\mu\Omega\sin^{2}\theta}{cR^{2}})
=\frac{\partial}{\partial\theta}[\frac{2\mu\Omega}{3cR^{2}}
P_{2}(\cos\theta)]
    \end{eqnarray}
where $P_{2}$ is the Lagendre polynomial of second degree.

If the exterior region is vacuum then
    \begin{eqnarray}
{\bf E^{out}}=-\nabla\Phi
      \end{eqnarray}
where
    \begin{eqnarray}
\nabla^2\Phi=0
    \end{eqnarray}

In order to solve the boundary condition (20) at $r=R$ the
solution of equation (22) must be \cite{7,8}
      \begin{eqnarray}
\Phi=-\frac{2\mu\Omega R^{2}}{3cr^{3}}P_{2}(\cos\theta)
    \end{eqnarray}
and hence
     \begin{eqnarray}
{\bf E}=-\frac{\mu\Omega R^{2}}{cr^{4}}[(3\cos^{2}\theta-1),
 2\sin\theta\cos\theta,0]
      \end{eqnarray}

Therefore one obtains the $r$ component of the induced quadrapole
electric field at the pole for the Newtonian case as
      \begin{eqnarray}
 E_{r}=-\frac{2\mu\Omega}{cR^{2}}
      \end{eqnarray}

Equating equations (16) and (25) one obtains
     \begin{eqnarray}
a=\frac{2\Omega R^{2}}{3c}
        \end{eqnarray}
Hence,
       \begin{eqnarray}
J/\Omega=\frac{2}{3}MR^{2}
       \end{eqnarray}
which in Newtonian mechanics, is regarded as the moment of
inertia of a hollow sphere. Hence $a$ in this case can be
regarded as the angular momentum per unit mass of a hollow
sphere.

In the above discussions the parameter $a$ plays the role in
determining the induced electric field only and since it has
been taken arbitrarily small it does not give rise to any
complicity in the spacetime geometry. Also the electromagnetic
fields are specialized to the lowest order expression in terms
of spherical harmonics. Therefore the spacetime geometry is
determined solely by the parameter $M$ i.e., the total mass of the
object. However one could start with any arbitrary value of $a$
but in that case departure from sphericity would prohibit in
obtaining a proper physical interpretation of the parameter $a$
in the language of Newtonian mechanics.

Nevertheless, since $a$ is a fixed parameter in the Kerr metric
so just like $M$ its physical meaning remains unaltered if one
considers the most general case.  Hence the parameter $a$ in the
Kerr metric may be interpreted as the angular momentum per unit
mass of a hollow object atleast within the above approximations.

Thus, in conclusion I would like to emphasize that the above result
though obtained from electromagnetic consideration, clearly indicates
that it is very much unlikely to have a physical source for the Kerr
field.

I am thankful to C. V. Vishveshwara for a critical reading of the
manuscript and for discussions.

\end{document}